\documentclass[fleqn,12pt,twoside]{article}

\usepackage[headings]{espcrc1}

\readRCS
$Id: espcrc1.tex,v 1.2 2004/02/24 11:22:11 spepping Exp $
\usepackage{graphicx}

\usepackage[figuresright]{rotating}

\newcommand{\AmS}{{\protect\the\textfont2
  A\kern-.1667em\lower.5ex\hbox{M}\kern-.125emS}}

\newcommand {\bea}{\begin{eqnarray}}
\newcommand {\eea}{\end{eqnarray}}
\newcommand {\be}{\begin{equation}}
\newcommand {\ee}{\end{equation}}

\title{Effects of the neutrino electromagnetic form factors on the neutrino and antineutrino mean free paths in dense matter}

\author{P.T.P. Hutauruk\address[parada]{Departemen Fisika, FMIPA, Universitas Indonesia, 
        Depok, 16424, Indonesia}, {A. Sulaksono\addressmark} and T. Mart\addressmark}

\runtitle{Effects of the neutrino electromagnetic form factors}
\runauthor{P. T. P. Hutauruk et al.}

\begin{document}
\maketitle

\begin{abstract}
We have studied the effects of the nucleon weak magnetism and the neutrino 
electromagnetic properties on the neutrino and antineutrino mean free paths 
for all types of neutrinos. In the calculation, we consider matters with and 
without neutrino trapping as the target. We have found that the difference 
between the two mean free paths depends not only on the neutrino energy, but
also on the matter density as well as matter constituents. Compared to the nucleon
weak magnetism, the neutrino electromagnetic properties are found to have 
negligible effects on this difference.   
\end{abstract}

\section{Introduction}
In the standard model, neutrinos have zero mass, magnetic moment, as well as 
electronic charge. However, there have been experimental upper 
limits on the electron-, muon- and tau-neutrino dipole moments given by 
${\mu}_{\nu_{e}} < 1.5\times 10^{-10} {\mu}_{B}$ \cite{Beac99} ,  
${\mu}_{\nu_{\mu}} < 7.4\times 10^{-10} {\mu}_{B}$ \cite{Krak90} and  
${\mu}_{\nu_{\tau}} < 5.4\times 10^{-7}{\mu}_{B}$ \cite{Coop92},  
where ${\mu}_{B}$ is the Bohr magneton. While neutrino charge radii of  
$R = (0.54 \pm 0.82)\times 10^{-6}~{\rm MeV}^{-1}$  (for electron-neutrino) and 
$R <0.40 \times 10^{-6}~{\rm MeV}^{-1}$ \cite{Vila95} (for muon-neutrino) have 
been experimentally evident, no experimental bound has been found so far on 
the charge radius of tau-neutrino \cite{Josh01}. On the other hand, the 
corresponding bounds from Super-K and SNO observations \cite{Josh01} are 
substantially larger, i.e., ${\mu}_{\nu_{\tau}}, {\mu}_{\nu_{\mu}} 
< 6.73 ~(5.77)\times 10^{-10} {\mu}_{B}$ and ${\mu}_{\nu_{e}} < 6.45 ~(5.65) 
\times 10^{-10} {\mu}_{B}$, whereas for the neutrino charge radii they 
obtained that $R_{\nu_{\mu}}, R_{\nu_{\tau}} < 2.31~(1.98) \times 
10^{-6}~{\rm MeV}^{-1}$ and $R_{\nu_{e}} <1.33~(1.16) \times 
10^{-6}~{\rm MeV}^{-1}$. Moreover, various astrophysical observations 
provide a limit on the neutrino magnetic moment in the range of  
$(1-4) \times 10^{-10} {\mu}_{B}$ \cite{Hagi02}, while from the plasmon 
decay in globular-cluster stars it is found that ${\mu}_{\nu} < 3\times 
10^{-12} {\mu}_{B}$  and $e_{\nu} < 2\times 10^{-14}e$ \cite{Raff99},
where $e$ is the proton charge.

Using the standard neutrino properties, Horowitz and Garc\'ia \cite{Horo05}
found that the muon-antineutrino mean free path in high density matter was 
considerably larger than the muon-neutrino one, provided that the weak magnetism  
of the nucleons is included. The sensitivity of differential cross section 
of the interaction between neutrinos and dense matter to the possibly nonzero 
neutrino electromagnetic properties has been also investigated \cite{Will05,Sul06}. 
It is found that the effects of the neutrino electromagnetic properties on 
the differential cross section become more significant for the neutrino 
magnetic moment ${\mu}_{\nu} > 10^{-10} {\mu}_{B}$ and the neutrino 
charge radius $R> 10^{-5}~{\rm MeV}^{-1}$. Motivated by this fact, 
in this paper we present the effect of the neutrino electromagnetic
properties on the difference between neutrino and its 
antineutrino mean free paths. To this end, we follow the same procedure as in 
Ref.~\cite{Sul06}, i.e., the differential cross section is calculated 
in a linear response with zero temperature approximation, while 
the leptons are assumed to be Fermi gas. In modeling the interacting 
nucleons, the relativistic mean field model inspired by 
 effective field theory \cite{Furu96} has been used to describe the 
non strange dense matter with and without neutrino trapping. 

\section{Neutrino and antineutrino mean free paths}
In this section we briefly discuss the analytic expression for the difference between neutrino-matter and antineutrino-matter  cross sections. In this case the electromagnetic form factors of the neutrino-electron and the weak magnetism of the nucleons are taken into account. We start with the Lagrangian density of the neutrino-matter interactions for each constituent in the form of 
\bea
{\mathcal L}_{\rm int}^{ j}&=&\frac{G_F}{\sqrt{2}}
\left(\bar{\nu}\Gamma_{\rm W}^{\mu} \nu\right)\left(\bar{\psi} J_{\mu}^{{\rm W}~{ j}} \psi\right)
+\frac{4 \pi \alpha}{q^2}\left(\bar{\nu}\Gamma_{\rm EM}^{\mu} \nu\right)\left(\bar{\psi} J_{\mu}^{{\rm EM}~{ j}} \psi\right),
\label{eq:lagden1}
\eea
where $G_F$ and $\alpha$ are the coupling constant of the weak interaction and the electromagnetic fine structure constant, respectively, and $j=n, p, e^{-}, {\mu}^{-}$. The parity violating vertex of neutrinos is given by $\Gamma_{\rm W}^{\mu}=\gamma^\mu(1-\gamma^5)$, while the electromagnetic properties of Dirac neutrinos are described in 
terms of four form factors, i.e., $f_{1\nu}, g_{1\nu}, f_{2\nu}$ and 
$g_{2\nu}$, which stand for the Dirac, anapole, magnetic, and electric 
form factors, respectively. The electromagnetic vertex $\Gamma_{\rm EM}^{\mu}$ contains electromagnetic form factors~\cite{Kerimov,Mourao}. Explicitly, it reads
\bea
\Gamma_{\rm EM}^{\mu}=f_{m\nu}\gamma^{\mu}
+ g_{1\nu}\gamma^{\mu}\gamma^{5}-(f_{2\nu}+ig_{2\nu}\gamma^5)
\frac{P^\mu}{2m_e},
\eea
where $f_{m\nu}=f_{1\nu}+(m_\nu / m_e)f_{2\nu}$, 
$P^{\mu} = k^{\mu} +k^{\mu\prime}$, $m_\nu$ and $m_e$ are the neutrino and 
electron masses, respectively.
In the static limit, the reduced Dirac form factor $f_{1\nu}$ and the 
neutrino anapole form factor $g_{1\nu}$ are related to the vector and 
axial-vector charge radii $\langle{R^2_V}\rangle$ and $\langle{R^2_A}\rangle$ 
through~\cite{Kerimov} 
\begin{equation}
f_{1\nu}(q^2)={\textstyle \frac{1}{6}} \langle{R^2_V}\rangle q^2 \quad\textrm{and}\quad 
g_{1\nu}(q^2)={\textstyle \frac{1}{6}}\langle{R^2_A}\rangle q^2,
\label{effac}
\end{equation}
where the neutrino charge radius is defined by $R^2$ = $\langle{R^2_V}\rangle$ + $\langle{R^2_A}\rangle$. In the limit of $q^2\to0$, $f_{2\nu}$ and $g_{2\nu}$  define
the neutrino magnetic moment  and the (CP violating) electric dipole moment, 
respectively~\cite{Kerimov,nardi}, i.e.,
\begin{equation}
\mu_{\nu}^{m}=f_{2\nu}(0)\mu_B \quad\textrm{and}\quad \mu_{\nu}^e=g_{2\nu}(0)\mu_B,
\label{mffac}
\end{equation}
where  $\mu_{\nu}^2$=${(\mu_{\nu}^m)}^{2}$+ ${(\mu_{\nu}^e)}^{2}$. The explicit forms of $J_\mu^{{\rm W}~j}$~\cite{Horo05} and  $J_\mu^{{\rm EM}~{ j}}$~\cite{Vogel} are given by
\be
J_\mu^{{\rm W}~j}=F_1^{{\rm W}~j}\gamma_\mu-G_A^{j}\gamma_\mu \gamma^5
+i F_2^{{\rm W}~j} ~\frac{\sigma_{\mu \nu}q^{\nu}}{2M},~ ~ ~ ~J_\mu^{{\rm EM}~{ j}}=F_1^{{\rm EM}~{ j}}\gamma_\mu+i F_2^{{\rm EM}~{ j}} ~\frac{\sigma_{\mu \nu}q^{\nu}}{2M}.
\ee
 
For antineutrinos, we must replace $G_A^{j}$ with $-G_A^{j}$.  In the limit of the photon point $q^2\to0$, for each type of neutrinos, the weak form factors $F_1^{\rm W}$, $G_A$ and $F_2^{\rm W}$ are given in Table~\ref{tab:copconst0}, whereas the electromagnetic form factors for each target  $F_1^{\rm EM}$ and $F_2^{\rm EM}$ are shown in Table~\ref{tab:copconst1}.

\begin{table}[htb]
\caption{Weak form factors in the limit  of $q^2\to0$. Here we use $\sin^2\theta_{\rm w}=0.231$, $g_A= 1.260$, $\mu_p=1.793$ and $\mu_n=-1.913$~\cite{Horo05}. The index $i$ indicates $e$, $\mu$ and $\tau$.}
\label{tab:copconst0}
\newcommand{\m}{\hphantom{$-$}}
\newcommand{\cc}[1]{\multicolumn{1}{c}{#1}}
\renewcommand{\tabcolsep}{1.45pc} 
\renewcommand{\arraystretch}{1.2} 
\begin{tabular}{@{}lrrc}
\hline Reaction & $F_1^{\rm W}$~~~~~~~ & $G_A$~ & $F_2^{\rm W}$ \\\hline
$\nu_i~n \rightarrow\nu_i~n $     &$-0.5$~~~~~~~  & $- g_A/2$  &$-(\mu_p-\mu_n)/2-2\sin^2\theta_{\rm w}\mu_n$\\
$\nu_i~p\rightarrow\nu_i~p$     &$0.5-2\sin^2\theta_{\rm w}$  & $g_A/2$ &~~$(\mu_p-\mu_n)/2-2\sin^2\theta_{\rm w}\mu_p$\\
$\nu_e ~e\rightarrow \nu_e~e$    &$0.5+2\sin^2\theta_{\rm w}$  & 1/2 &0\\
$\nu_\mu~\mu\rightarrow\nu_\mu~\mu $     &$0.5+2\sin^2\theta_{\rm w}$  & 1/2 &0\\
$\nu_{\mu,\tau}~e\rightarrow\nu_{\mu,\tau}~e $ &$-0.5+2\sin^2\theta_{\rm w}$  & $-1/2$  &0 \\
$\nu_{e,\tau}~\mu\rightarrow\nu_{e,\tau}~\mu $ &$-0.5+2\sin^2\theta_{\rm w}$  & $-1/2$  &0 \\\hline
\end{tabular}\\[2pt]
\end{table}

\begin{table}[htb]
\caption{Electromagnetic form factors of neutrinos in the limit  of $q^2\to0$~\cite{Vogel}.}
\label{tab:copconst1}
\newcommand{\m}{\hphantom{$-$}}
\newcommand{\cc}[1]{\multicolumn{1}{c}{#1}}
\renewcommand{\tabcolsep}{4.5pc} 
\renewcommand{\arraystretch}{1.2} 
\begin{tabular}{@{}ccc}
\hline
 Target &$F_1^{\rm EM}$ & $F_2^{\rm EM}$ \\\hline
$n$     &0              & $\mu_n$   \\
$p$     &1              &  $\mu_p$  \\
$e$     &1              & 0   \\
$\mu$   &1               & 0 \\
\hline
\end{tabular}\\[2pt]
\end{table}
Using the Lagrangian density given by Eq.\,(\ref{eq:lagden1}), we can obtain 
the neutrino-matter and antineutrino-matter differential cross sections~\cite{Sul06}.
Their difference ($\Delta \sigma$) for each type of neutrino can be calculated from
\begin{eqnarray}
\Biggr[\frac{1}{V}\frac{d^3{(\Delta\sigma)}}{d^2{\Omega}d^2E_{\nu}^{'}}\Biggr]_i&=&\sum_{j=p,n,e^{-},\mu^{-}} \frac{1}{4\pi^{2}}\frac{E_{\nu}^{'}}{E_{\nu}} \Biggr[\Biggr(\frac{G_F}{\sqrt{2}}\Biggr)^{2}{ (2E-q_0)\Biggr(F_1^{Wj}G_A^{j} 
+ \frac{m}{M}F_2^{Wj}G_A^{j}\Biggr)\Pi_{VA}^{j}} \nonumber \\
&+& \frac{8 G_F{\pi}{\alpha}}{3 \sqrt{2}}{q^2 R ^2 (2E-q_0) \Biggr(\frac{m}{M}F_2^{EMj}G_A^{j} 
+ F_1^{EMj}G_A^{j}\Biggr)\Pi_{VA}^{j}}\Biggr],
\label{delcc}
\end{eqnarray}
where $E_{{\nu}}$ and $ E'_{{\nu}}$ are the initial and final neutrino energies, 
respectively, while $M$ is the target mass.  For the nucleon $m$ is the effective 
mass $M^*$, whereas for the lepton $m$=$M$. In Eq.~(\ref{delcc}) the charge 
radius of the neutrino is indicated by $R$. The explicit values of the target form 
factors $F_i$ and $G_A$ for each reaction are listed in Tables~\ref{tab:copconst0} 
and \ref{tab:copconst1}. The axial-vector polarization 
tensor $\Pi_{VA}$ is given by 
\begin{eqnarray}
\Pi_{VA}=\frac{iq^2}{8\pi\vert\,\vec{q}\,\vert^3}\left[(E^2_F-E^{* 2}) +q_0 (E_F-E^*)\right], 
\label{PiVA}
\end{eqnarray}
where $E_F$ and $E^{*}$ denote the Fermi and effective energies of each target,
respectively. 
From Eq.\,(\ref{delcc}) it is obvious that, qualitatively, the  charge radius 
of the neutrino yields some correction to the cross section difference, whereas 
this is not the case for the neutrino dipole moment. Its manifestation in the 
form of quantitative difference in the mean free path will be discussed in the 
following section.   

\section{Results and discussions}
\begin{figure}[t]
\begin{minipage}[t]{75mm}
\includegraphics[width=20pc, height=20pc] {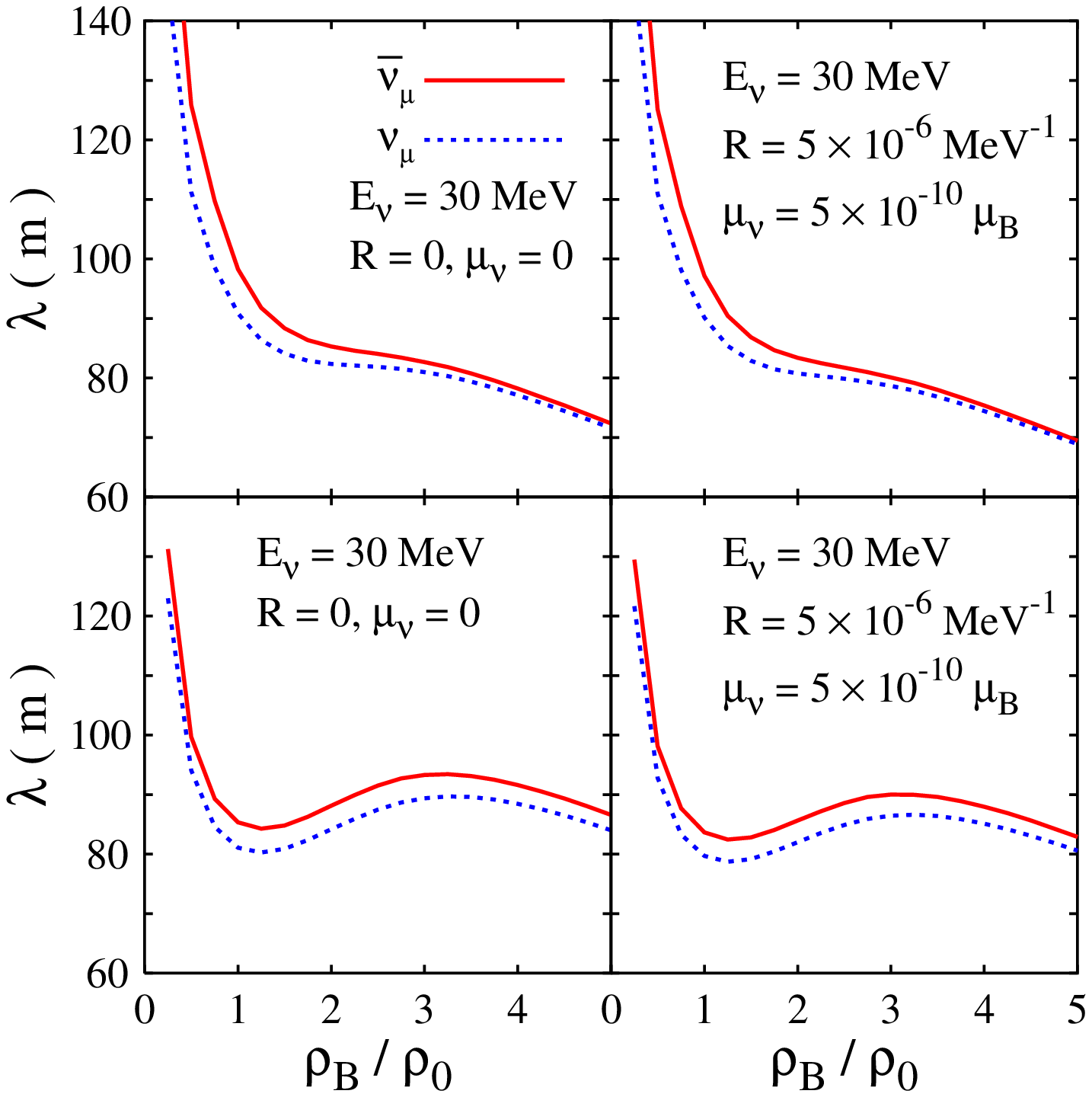}
\caption{The muon-neutrino and muon-antineutrino mean free paths as a function 
  of the ratio between nucleon and nuclear saturation densities. Results 
  for the neutrinoless matter are shown in the upper panels, whereas results for 
  the neutrino trapping with $Y_{le}$= $0.3$ are shown in the lower panels. }
\label{muon1}
\end{minipage}
\hspace{5mm}
\begin{minipage}[t]{75mm}
\includegraphics[width=20pc, height=20pc] {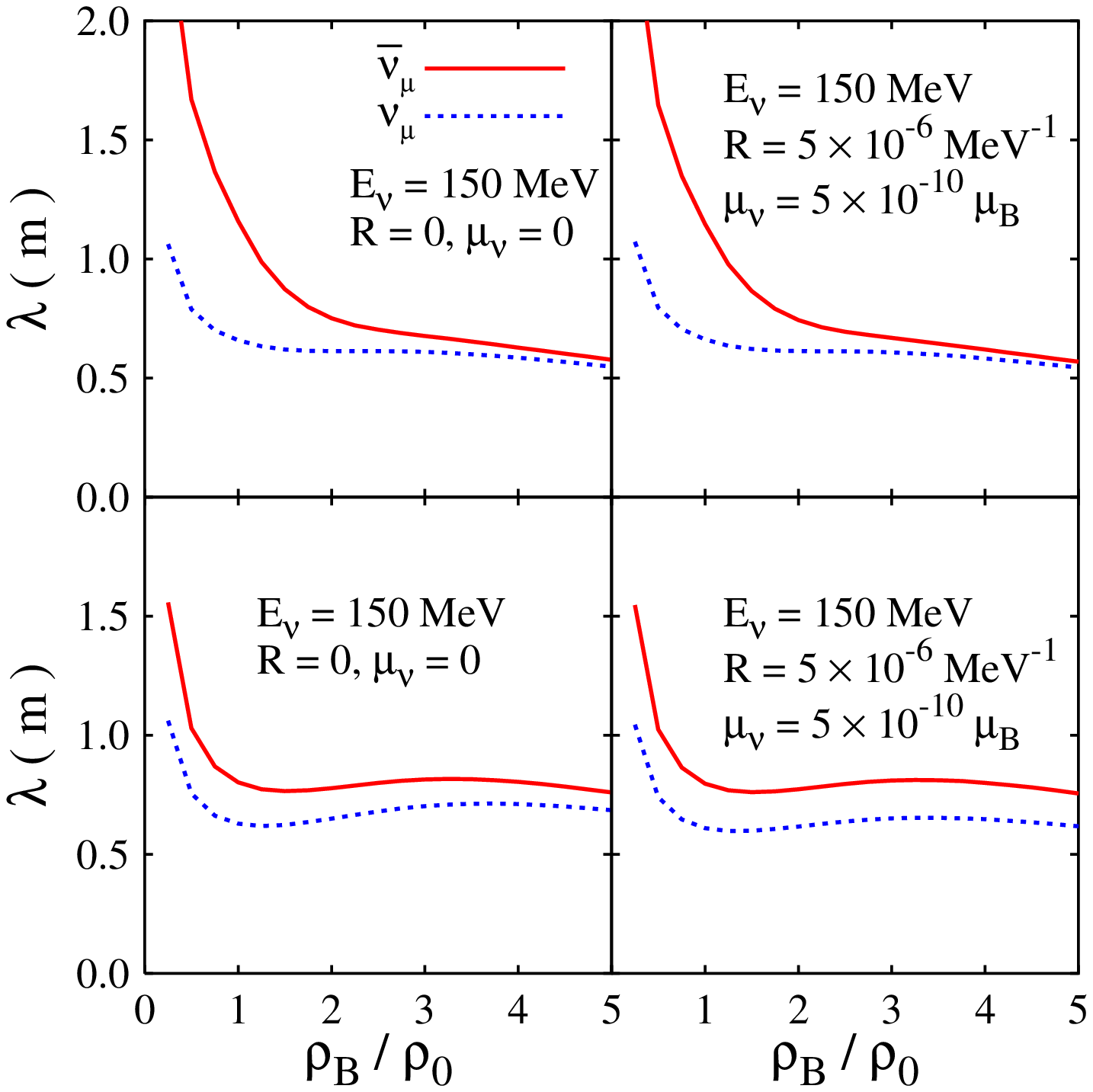}
\caption{Same as Fig.~\ref{muon1}, except with the muon-neutrino energy 
$E_{\nu}$= $150$ MeV.}
\label{muon2}
\end{minipage}
\end{figure}
\begin{figure}[t]
\begin{minipage}[t]{75mm}
\includegraphics[width=80mm] {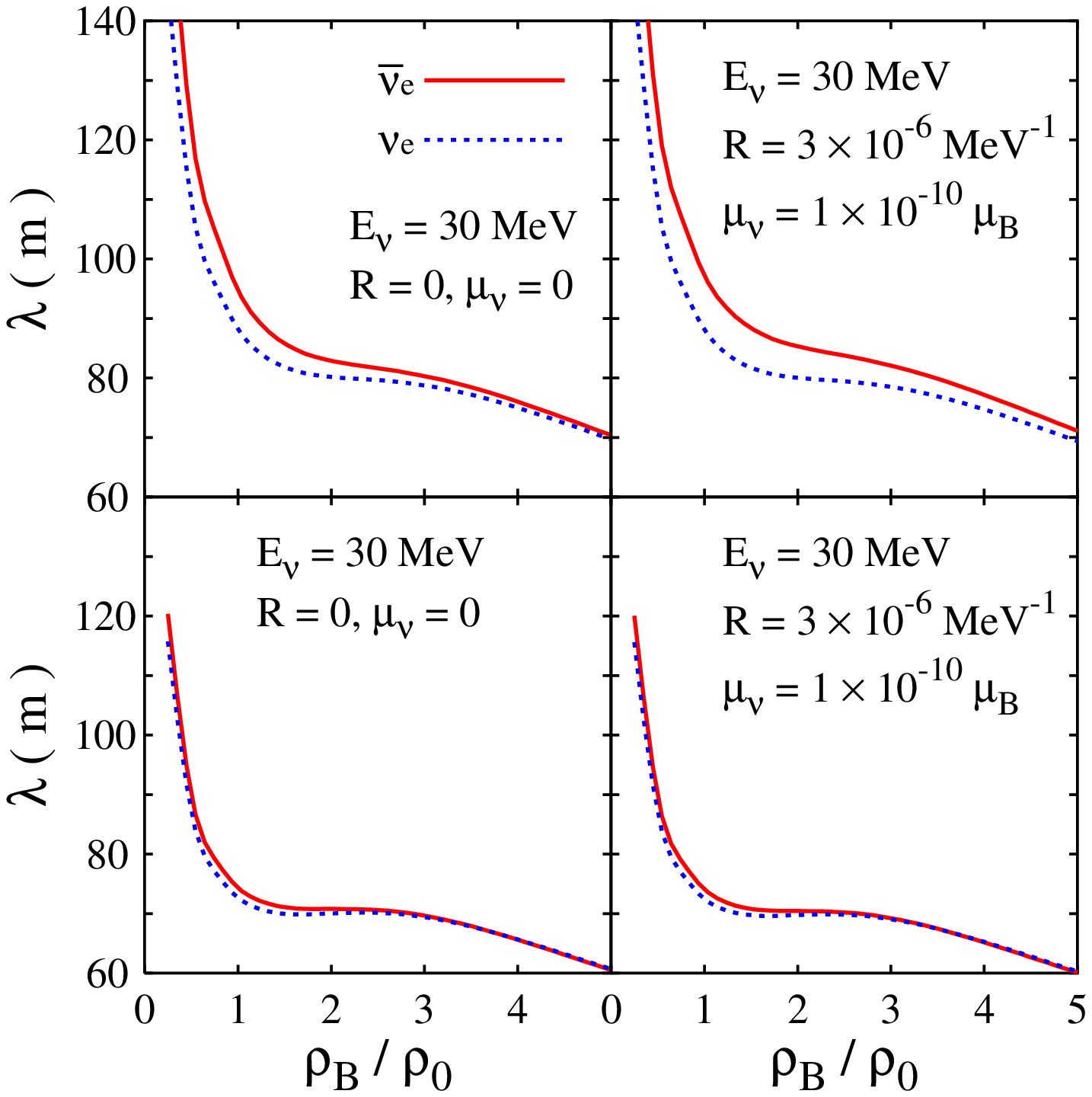}
\caption{Same as Fig.~\ref{muon1}, except for the electron-neutrino case.}
\label{elec}
\end{minipage}
\hspace{2mm}
\begin{minipage}[t]{75mm}
\includegraphics[width=80mm] {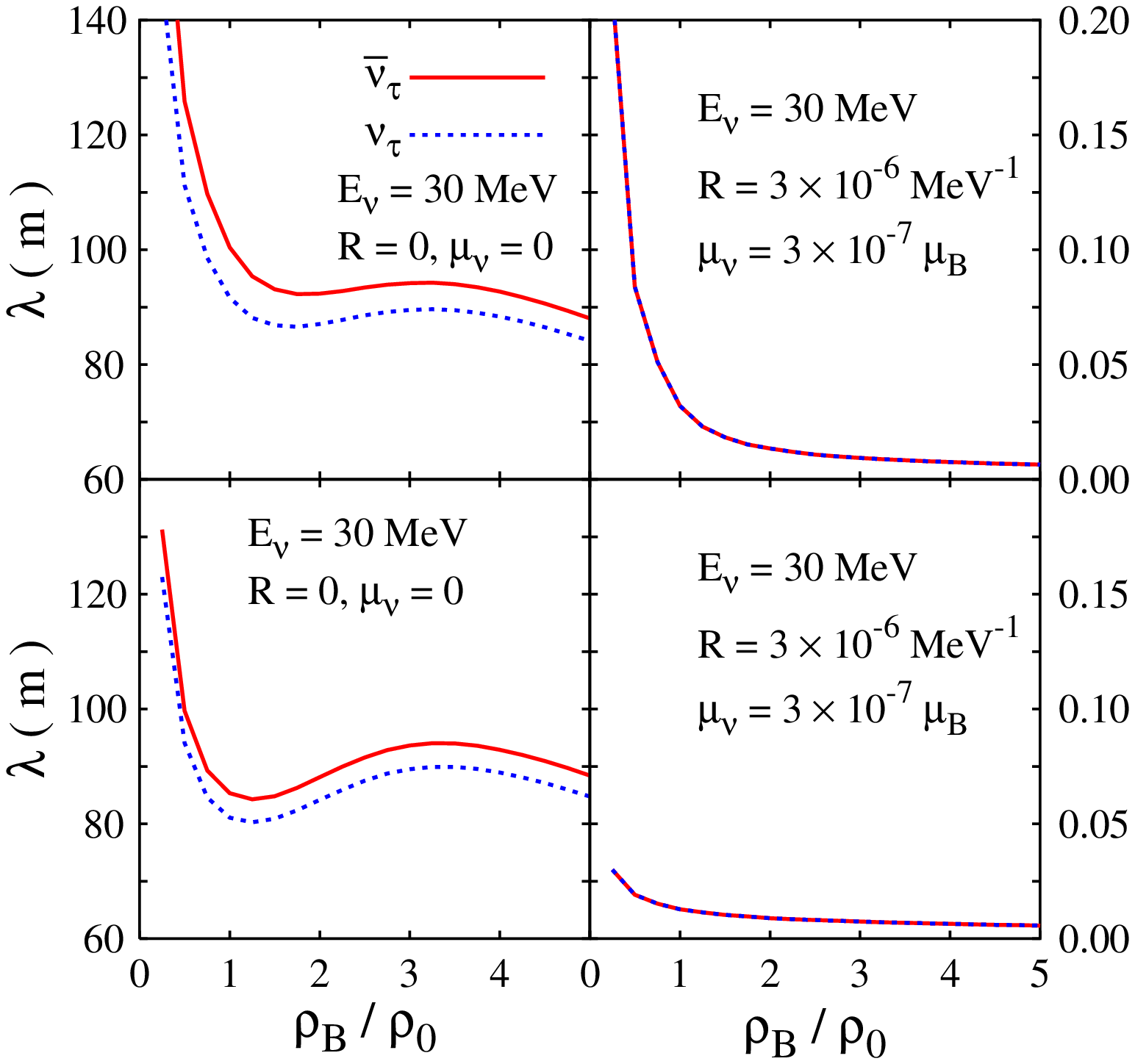}
\caption{Same as Fig.~\ref{muon1}, except for the tau-neutrino case.}
\label{tau}
\end{minipage}
\end{figure}

The neutrino and antineutrino mean free paths ($\lambda_{\nu,~\bar{\nu}}$) 
as a function of the density at a certain neutrino energy is obtained by 
integrating the corresponding cross section over the time and vector 
components of the neutrino momentum transfer~\cite{Horo05}. The 
$\lambda_{\nu}$ and $\lambda_{\bar{\nu}}$ for the case where the projectiles 
are  electron-, muon-, tau-neutrinos, as well as their antineutrinos, are 
shown in Figs.~\ref{muon1},~\ref{muon2},~\ref{elec} and ~\ref{tau}. In this 
calculation we consider two types of matters, i.e., matters with and without 
neutrino trapping. In the latter, it is dominated by neutrons and followed 
by a small number of protons, electrons, and muons, which start to emerge at 
relatively large densities. 

The existence of the neutrinos 
in matter also allows for the presence of a relatively large number of protons 
and electrons as compared to the case of neutrinoless matter. The appearance 
of these constituents is then followed by the appearance  of  a small number 
of muons at a density larger than twice of the nuclear saturation density. 
The relative fraction of the individual constituent of matter as a function 
of the ratio between  nucleon and nuclear saturation densities can be found 
in Ref.~\cite{Sul06}. In order to see the effect more clearly, in this 
calculation we use the relatively large values of neutrino dipole 
moments  and charge radii as compared to the stringent bounds. 

In Fig.~\ref{muon1}, it can be seen that in the case of matter without 
neutrino trapping,  contribution of the neutrino charge radius is much 
smaller than that of the nucleon weak magnetism. Therefore, it does not 
give a visible effect in the difference between muon-neutrino and its 
antineutrino mean free paths. On the other hand, in the case  of 
neutrino trapping,  but with  zero muon-neutrino dipole moment and 
charge radius, the difference is quite pronounced, especially at high 
densities. Albeit similar with the neutrinoless matter case, contribution 
from the neutrino charge radius is very small, and therefore it can not 
be observed within this kinematics. It is also apparent that in both 
cases of matter with densities around $(2-3)\rho_0$, $\lambda_{\nu}$ 
and $\lambda_{\bar{\nu}}$ behave differently, i.e., if neutrinos are 
present in the matter, the mean free paths increase with respect to 
the density, but if  neutrinos are not present, the opposite phenomenon 
is observed. 

For the case of neutrinoless matter, if the muon-neutrino 
energy is increased to 150 MeV (Fig.~\ref{muon2}), the effect of 
the charge radius for this neutrino type can be neglected and the mean 
free path difference significantly increases at low densities. On the 
contrary, in the case of neutrino trapping, the mean free path 
difference appears to be more or less constant  for all densities 
and a substantial enhancement caused by the neutrino charge radius 
contribution shows up only at high densities. As shown in Fig.~\ref{elec}, 
in spite of the fact that the effect is less significant in neutrinoless 
matter at high densities, contribution from the neutrino charge radius 
yields an enhancement to the difference between $\lambda_{\nu}$ and 
$\lambda_{\bar{\nu}}$. Interestingly, for the case of zero neutrino 
dipole moment and charge radius, but with neutrino trapping, the mean 
free path difference is suppressed. Thus, for this kind of matter, the effect 
of the neutrino charge radius in the mean free path difference is 
insignificant. 

Similar to the muon-neutrino case, a different behavior of the mean free 
path is also observed in the tau-neutrino case at the densities around 
$(2-3) \rho_0$ (see Fig.~\ref{tau}). 
Due to the relatively large value of the  tau-neutrino dipole 
moment used in this calculation, contribution from the protons to both 
mean free paths turns out to be very large and, as a consequence, the difference 
between $\lambda_{\nu}$ and $\lambda_{\bar{\nu}}$  is almost negligible.

\section{Conclusions}
We have calculated the neutrino and antineutrino mean 
free paths for all types of neutrinos and study their differences 
with various mechanisms, i.e., by using different types of matters 
(with and without neutrino trapping) as well as by including the weak 
magnetism of the nucleon and the electromagnetic form factors of the 
neutrinos. We have found that the difference in the neutrino and 
antineutrino mean free paths depends not only on the matter density, 
but also on the neutrino energy and the matter constituents. 
However, variation of the latter affects the mean free path  
difference in a distinct way. The effect of the neutrino electromagnetic 
form factors has been also studied. It is found that the corresponding 
contribution can be neglected as compared to the contribution from the 
nucleon weak magnetism.    

\section*{Acknowledgment}
AS and TM acknowledge the supports from the Faculty of Mathematics and 
Sciences, University of Indonesia, and from the Hibah Pascasarjana grant.


\begin{thebibliography}{20}
\bibitem{Beac99} J.F. Beacom and P. Vogel, Phys. Rev. Lett.  89 (1999) 5222.
\bibitem{Krak90} D.A. Krakauer  {\sl et al}., Phys. Lett.  B 252 (1990) 177.
\bibitem{Coop92} A.M Cooper-Sarkar  {\sl et al}., Phys. Lett.  B 280 (1992) 153.
\bibitem{Vila95}   P. Vilain {\sl et al.}, Phys. Lett.  B 345 (1995) 115.
\bibitem{Josh01} A.S. Joshipura and S Mohanty, hep-ph/0108018 (2001).
\bibitem{Hagi02} Particle Data Group,  K. Hagiwara  {\sl et al}., Phys. Rev. D 66 (2002) 010001.
\bibitem{Raff99} G.G. Raffelt, Phys. Rep. 320 (1999) 319.
\bibitem{Horo05} C.J. Horowitz and M.A. Per\'ez-Garc\'ia, Phys. Rev. C 68 (2003) 025803.
\bibitem{Will05} C.K. Williams, P.T.P. Hutauruk, A. Sulaksono and T. Mart, Phys. Rev. D 71 (2005) 017303.
\bibitem{Sul06} A. Sulaksono, C.K. Williams, P.T.P. Hutauruk, and T. Mart, Phys. Rev. C 73 (2006) 025803.
\bibitem{Furu96} R.J. Furnstahl, B.D. Serot and H.B. Tang, Nucl. Phys.  A 598 (1996) 539; 
Nucl. Phys. A 615 (1997) 441.
\bibitem{Kerimov} B.K. Kerimov, M. Ya Safin and H. Nazih, Izv. Akad. Nauk. SSSR Ser. Fiz. 52 (1998) 126.
\bibitem{Mourao} A.M. Mour\~{a}o, J. Pulido, and J.P. Ralston, Phys. Lett B 285 (1992) 364.
\bibitem{nardi} E. Nardi, AIP Conf.\ Proc.\  {670} (2003) 118.
\bibitem{Vogel} P. Vogel and J. Engel, Phys. Rev. D 39 (1989) 3378.
\end{thebibliography}
\end{document}